\documentstyle[aps,prl,rotate,epsf,epsfig,amssymb,floats]{revtex}

\def \be   { \begin{displaymath} }
\def \ee   { \end{displaymath}   }       
       
\def \ben  { \begin{equation}    }
\def \een  { \end{equation}      }       

\def \bea  { \begin{eqnarray*}   }       
\def \eea  { \end{eqnarray*}     }

\def \bean { \begin{eqnarray}    }       
\def \eean { \end{eqnarray}      }

\def \bfe   { {\mathbf e } }               

\def \Ref#1{(\ref{#1})}
\def \fie   { \varphi      }
\def \eps   { \varepsilon  }

\begin{document}
\twocolumn[\hsize\textwidth\columnwidth\hsize\csname
@twocolumnfalse\endcsname


\title{Modeling Disordered Quantum-Systems with Dynamical Networks}

\author{Rochus Klesse\cite{r_address} and Marcus Metzler\cite{m_address}}

\address{Institut f\"ur Theoretische Physik, Universit\"at zu
  K\"oln, D-50937 K\"oln, Germany}  
\date{\today}

\maketitle

\begin{abstract}
It is the purpose of the present article to show that so-called network
models, originally designed to describe static properties of disordered 
electronic systems, can be easily generalized to quantum-{\em
dynamical} models, which then allow for an investigation of dynamical
and spectral aspects. This concept is exemplified by the
Chalker-Coddington model for the Quantum Hall effect and a
three-dimensional generalization of it. We simulate phase coherent
diffusion of wave packets and consider spatial and spectral
correlations of network eigenstates as well as the distribution of
(quasi-)energy levels. Apart from that it is demonstrated how network
models can be used to determine two-point conductances. Our numerical
calculations for the three-dimensional model at the Metal-Insulator
transition point delivers among others an anomalous diffusion
exponent of $\eta = 3 - D_2 = 1.7 \pm 0.1$. The methods presented here
in detail have been used partially in earlier work.
\end{abstract}

\vskip 2pc]

\section{Introduction}
\label{sec-intro}

Models formulated in scattering theoretical terms have been
proven to be convenient for the investigation of 
quantum particle- or classical wave-propagation in disordered media.  
Examples related to Anderson localization \cite{anderson} are the
scattering model of a one-dimensional (1D) disordered conductor used 
by Anderson et al.
\cite{anderson-thouless-abrahams-fisher}, it's three-dimensional (3D)
version introduced by Shapiro \cite{shapiro}, and the Chalker-Coddington  
network of the Quantum Hall transition \cite{chalker-coddington}.
Numerous generalizations and variations of these models have been used
in order to study various topics in different fields
\cite{general,edrei,lee,zirnbauer,klesse-metzler,chalker-dohmen}.

The common concept in these approaches is the scattering theoretical
formulation of the problem. This leads to models which can be described
as networks consisting of an array of local scattering centers, 
linked together by 1D channels \cite{shapiro}. 
Usually, the {\em static} properties of these models are 
investigated, like probabilities 
for scattering through the network or correlations of stationary 
scattering states.

It is the purpose of the present article to show that in general 
these models can be easily extended to {\em dynamical} network models,
which then allow to study dynamical as well as spectral aspects
of the respective problem under consideration. In particular, we consider
dynamical versions of the Chalker-Coddington model
and of a 3D generalization of it \cite{chalker-dohmen}.
Within these models we 
investigate numerically phase coherent diffusion of wave-packets,
whereby special attention is paid to anomalous diffusion near the
localization-delocalization transition point.
It is demonstrated how  spectral information can be 
extracted from these models, like (quasi-)energy spectra or the
local density of states. Moreover, we present an example
within the framework of a network model  for
determining conductances of an integer quantum Hall system contacted
by point-contacts.

A network model is mathematically formulated by an explicitly
defined unitary matrix, the network operator $U$, that determines 
the scattering processes between internal network channels. 
By definition, in the dynamical model this operator $U$ will
describe the evolution of network-states $\Psi$ over a microscopic
time interval $\tau$, $\Psi \stackrel{\tau}{\longrightarrow}
U\Psi$. By this, a discrete time evolution is given: $\Psi(k\tau) =
U^k \Psi(0)$, where $k=0,1,2,\dots$. Further, network eigenstates $\Phi_n$
are defined as eigenvectors of $U$. The phases $\omega_n \in [0,2\pi[$ of the
unimodular eigenvalues $\exp(i\omega_n)$ will be interpreted as
quasi-energy levels.

The main part of this work consists of numerical simulations within
the two aforementioned models in order to demonstrate that the
concept of dynamical networks is self-consistent and, above all, that
it delivers results consistent with previous results and general
theory of disordered systems. These simulations also clearly show that 
network models are convenient for numerical purposes.
They allow e.g. to calculate efficiently the (fully phase coherent)
time evolution 
of wave-packets diffusing in comparably large systems. 
Furthermore, the particle energy (to be distinguished
from the network specific quasi-energy), becomes simply a parameter of
the model. Due to this it is possible to focus at specific and sharply
defined energy regimes, which will turn out to be
advantageous when investigating critical level statistics. 

Our approach is related to that of Edrei et al. \cite{edrei}, where
a network model has been used for calculating wave propagation through
random media. Actually, the definition of network states and operator
used here are, in principle, identical to those in \cite{edrei}.
However, their work concentrated on stationary states of networks
with open boundaries in order to determine transmission coefficients.
In contrast to that, here we study the time evolution explicitly and
consider in particular closed systems. The latter offers the
opportunity to define and to investigate correlations 
of network eigenstates and quasi-energy levels. 

Viewed as a time evolution operator, the network operator
is also related to so called quantum maps or Floquet operators,
which attracted recently considerable attention in quantum
chaology \cite{smilansky}. It seems to be that dynamical networks
and quantum maps are identical concepts, applied in different 
physical contexts: the former to diffusive, the latter to chaotic
systems. 
However, the precise relation between these two classes of systems is
not entirely clear to us yet and deserves further
investigation in the future.

The concepts presented here have been applied partially in previous work,
e.g. for determining critical eigenstates \cite{klesse-metzler}, local
density of states \cite{huckestein-klesse}, critical level statistics
\cite{klesse-metzler2,metzler_varga,Metzler2} and quantum diffusion 
\cite{huckestein-klesse2}. Here we explain in more detail the numerical
methods used therein. However, also the reader not familiar with 
network models might find this paper worth
reading, since it deals with current topics in the field of 
critical disordered systems, like the multifractality of the local 
density of states, critical diffusion or energy level correlations.

The article is organized as follows: We start with a definition of
dynamical network models, which tries to be a compromise
between exactness and generality on the one hand and the amount of
formal effort on the other. Next is the introduction of the two
specific models mentioned above, within which several numerical
simulations will be performed.
Beginning with the simulation of diffusing wave packets we come to
the calculation of the local spectral density. After that we turn to
the distribution of network quasi-energy levels. The next section is
concerned with correlations in the local density of critical network
states, followed  by a section in which the network model is used for
the calculation of two-point conductances. We conclude with a summary
and general remarks.


\section{General Definitions}
\label{sec-network}
\begin{figure}[htb] 
\begin{center}
        \epsfxsize 8cm
        \epsffile{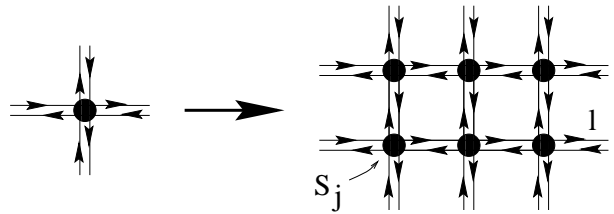}
\end{center}
\caption{The basic module of a network are elementary scatterers of the
  same type. 
}
\label{fig-scatterer}
\end{figure}

Consider a $d$-dimensional array of scattering centers, where each
of them has $n$ outgoing and incoming channels. Let them be labeled by
a lattice index $j$ and characterized by $n\times n$ unitary matrices
$S_j$. Then, connecting in a certain manner outgoing with incoming
channels of neighboring scatterers, one ends up with a network
(Fig. \ref{fig-scatterer}). We denote its scattering centers as {\em
nodes} and  internal channels as {\em links}, labeled by another 
index $l$. 

\begin{figure}[htb] 
\begin{center}
        \epsfxsize 8cm
        \epsffile{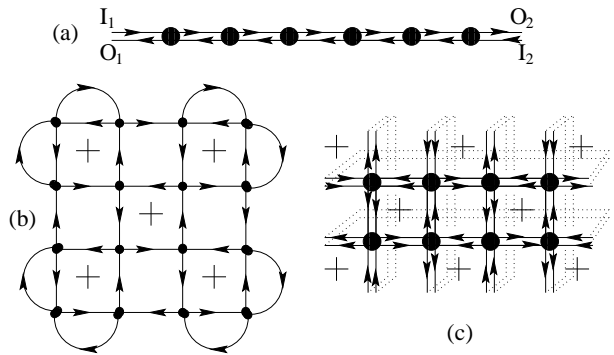}
\end{center}
\caption{Several networks of various dimensions and channel numbers $n$:
  (a) a 1D network with $n=2$ modeling a 1D chain of scatterers. (b)
  the two dimensional Chalker-Coddington network for the integer
  quantum Hall Effect, $n=2$. (c) the same as (b), but with spin
  degree of freedom, thus here $n=4$. The network (a) is open, (b)
  shows an example of a closed network with reflective boundaries
  whereas the closed network (c) has periodic boundary conditions.  
}
\label{fig-examples}
\end{figure}

A network is closed when all channels start from and end in network
nodes, otherwise it is open. In the latter case some channels start
from or end somewhere outside the network, forming input ($I_m$) and
output channels ($O_l$) of the network. Thereby, the number $N_c$ of
input channels always equals the number of output channels. Such an
open network can be viewed as a scatterer with a complex internal
structure and the transmission through it can be described by an $N_c$
dimensional matrix $\cal S$, which maps the input amplitudes in $I_m$
to the output channels $O_l$. Both types of networks are illustrated
by examples shown in Fig. \ref{fig-examples}.
 
The open network (a) would be for example appropriate for the study of
transmission or conductance \cite{edrei}, whereas the closed networks
(b) and (c) could be used e.g. for determining eigenstates, as we will
see in the following sections. 

One can choose reflecting or periodic boundaries for the network.
Reflecting boundaries are realized by bending the outgoing
channels of one side to neighbored incoming ones of the same side,
periodic boundaries by linking the outgoing channels to the incoming
ones of the opposite side (see Fig.~\ref{fig-examples}, (b) and (c) ).

A network state can be introduced as the set of complex amplitudes
$\psi_l$ for all links $l=1,\dots,N$, $\Psi = \{ \psi_l \}$. Per
definition, the 
network states form a vector space, spanned e.g. by the orthonormal
unit states $\bfe_{l'} = \{ \delta_{l'l} \}$. The scalar product is given
by $\langle \Phi, \Psi \rangle = \sum \phi^*_l \psi_l$.
When the states are interpreted as wave functions of a particle,
the squared amplitude $|\psi_l|^2 = |\langle \bfe_l , \Psi \rangle|^2$
of a normalized state denotes the probability to find the particle on
link $l$. 

In physically realized networks
only those states $\Psi$ are feasible, which have link amplitudes
$\psi_l$ obeying the scattering condition at every node of the
network. This means that for every outgoing link $l$ of a node $j$ with
a scattering matrix $S_j=\{t^j_{lm}\}$ the equation
\begin{equation}\label{scatteringcondition}
\psi_l = \sum_{m:l} t^j_{lm} \psi_m
\end{equation}
has to be fulfilled, whereby $m$ runs over the $n$ incoming links
of that scatterer $j$ from which link $l$ exits (indicated by the
symbol $m:l$).
Stationary or steady states of the network are states
satisfying this condition at each node. 

The right hand side of the stationarity condition
Eq. (\ref{scatteringcondition}) can be read as the action
of an operator $U$ on the state $\Psi = \{ \psi_l \}$, 
$$
 (U \Psi)_l = \sum_{k=1}^N U_{lk}\psi_k \equiv \sum_{m:l} t^j_{lm} \psi_m.
$$
Notice that the second sum contains only $n$ summands, and not $N$
(number of all links in the network) as in the first. This implies
that only $n$ of 
of the $N$ coefficients $\{U_{lk}\}_{k=1,\dots, N}$ are non-vanishing
(i.e. $U_{lk}$ is a {\em sparse} N-dimensional matrix).
For closed networks, this operator $U$ is unitary due to the unitarity
of the local scattering matrices $S_j$. Obviously, the stationarity
condition for a state $\Psi$ can then be written as 
\begin{equation}\label{upsigleichpsi}
U\Psi = \Psi.
\end{equation}
For open networks $U$ is no longer unitary, because in that case the
probability density conservation is violated by the leakage at the outgoing
channels. In this case $U$ can still be used to formulate the
stationarity condition, however, additional care has to be taken of the
open input and output channels. We will discuss this in detail for
one example in Sec.~\ref{sec-pconduct}.

\begin{figure}[htb] 
\begin{center}
        \epsfxsize 8cm
        \epsffile{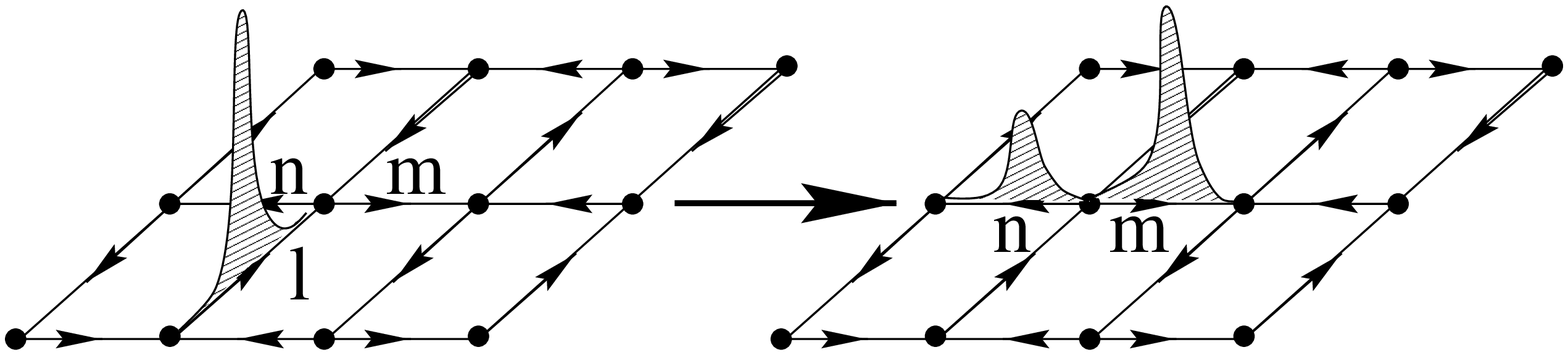}
\end{center}
\caption{ The action of $U$ on a basic state $\bfe_l$. It maps the
  incoming amplitude into the outgoing channels $m,n$ according to
  the transmission coefficients $t_{ml},t_{nl}$. The corresponding physical
  process can be thought of as the scattering of an incident wave packet
  into the outgoing channels after a characteristic time $\tau$.}
\label{fig-u}
\end{figure}

It is instructive to consider the action of $U$ on a basic state
$\bfe_l$,
\begin{equation}\label{udef}
U \bfe_l = \sum_{l:m} t^j_{ml} \bfe_m,
\end{equation}
where the sum runs over all outgoing links $m$ of scatterer $j$ in
which $l$ enters (denoted by $l:m$). Operator $U$ simply maps the
incoming amplitude  
to the outgoing links of the scatterer according to the transmission
coefficients $t^j_{lm}$ (Fig.~\ref{fig-u}).
A corresponding physical process can be thought of as the scattering of
an incident wave-packet at a node into the outgoing channels after a
characteristic time $\tau$. Therefore it is suggestive
to consider the action of $U$ on an arbitrary network state $\Psi$ as
its time evolution over a certain microscopic time interval $\tau$,
\ben\label{dynamics}
 \Psi(t_0 + \tau) \equiv U \Psi(t_0).
\een
By this, a dynamic for the network is defined by a stepwise time evolution
\begin{equation}\label{dynamic}
\Psi(t_0+ k\tau) = U^k \Psi(t_0)
\end{equation}
for an integer number of time steps $k= 0,\pm 1, \pm2, \dots$. (In the
following we often set $\tau \equiv 1$ and then write $\Psi(t) = U^t
\Psi(0)$.) Apparently, stationary states remain unchanged by these
dynamics, as it should be.

According to these dynamics, for closed networks a real spectrum
$\{\omega_n\}$ is given by the phases $\omega_n$ of the complex
eigenvalues of $U$,
\begin{equation}
U \Phi_n = e^{i\omega_n}\Phi_n.
\end{equation}
In order to avoid ambiguity we restrict the
phases or quasi-energies to $[0,2\pi[$. With respect to the dynamics
\Ref{dynamic} the corresponding eigenvectors $\Phi_n$ oscillate with
frequencies 
$\omega_n$, 
\begin{equation}
  \Phi_n(t) = U^t \Phi_n(0) = e^{i\omega t} \Phi_n(0).
\end{equation}
Additionally, we define a local density of network states
(LDOS) by
\begin{equation}
  \label{ldos}
  \rho_l(\omega) = \sum_n |\phi_{n,l}|^2 \delta(\omega-\omega_n).
\end{equation}

\section{Special Network Models}
\label{chap-models}
After we have introduced network 
models in general, let us now consider two special models in more detail:
the two-dimensional (2D) network introduced by Chalker and 
Coddington \cite{chalker-coddington} to describe Quantum-Hall systems and a
3D network very similar to the model proposed by
Chalker and Dohmen to describe multi-layer systems
\cite{chalker-dohmen}. Originally, both models are  purely static, but
they can be easily extended to dynamical models by providing them with
the discrete time evolution \Ref{dynamic}, as we will do here.

The Chalker-Coddington network is designed in the style of the
networks introduced by Anderson et
al. \cite{anderson-thouless-abrahams-fisher} and Shapiro
\cite{shapiro} 
in the early 1980s,  where a scattering formalism is used to 
describe Anderson localization. 
It provides a semi-classical description of a
2D electron in a strong perpendicular magnetic field $B$
and a potential $V$.
The random potential $V$ is assumed to be smooth
with a correlation length $a$ large
compared to the magnetic length $l_c$
\cite{tsukada,prange-joynt,trugman,fertig,iordansky,buch}.
Classically, the electron executes a fast
cyclotron motion on a circle of radius $l_c$, the center of which drifts
slowly along contours $V(r) = const.$. For $l_c \ll a$ it is
justified to separate cyclotron and drift motion.
Quantizing the first yields the Landau energies $E_n=
\hbar\omega(n + \frac{1}{2})$. The drift of the center coordinate can be
treated semi-classically by a WKB- or stationary-phase-approximation 
\cite{tsukada,prange-joynt,fertig}, leading to the following picture:
The probability density of an eigenstate at energy $E$ in the $n$th
Landau band is concentrated along contours $V(c_l) \equiv E-E_n$ of the
disorder potential. In local coordinates $u,v$ parallel ($u$)
and transverse ($v$) to a contour $l$ the wave function $\Psi_E(r)$
can be well approximated by
\begin{equation}
\label{contour-state}
  \Psi_E(u,v) = \psi_l \frac{1 }{ \sqrt{ v_d(u)}}  \varphi_n(v/l_c)e^{i\phi(u,v)} ,
\end{equation}
where $\varphi_n(x)$ denotes the $n$th Hermite polynomial and $\phi(u,v)$
a gauge dependent phase. Because of the normalization by the local
drift velocity $v_d(u)$, the squared modulus of the complex coefficient
$\psi_l$ gives the net current transported by the state along
the contour $l$ in direction of the drift motion. The approximation is
valid everywhere where the contour line is 
well separated from others by a distance large compared to
$l_c$.  At saddle points of the disorder potential lying at
energies close to $E-E_n$, such that the distance between two contour
lines becomes of order $l_c$, it breaks down.
However, at these points the 
tunneling between different contour lines 
can be described by local $2\times 2$ scattering matrices $S_j(E)= \{
t^j_{ml}(E)
\}$, relating the coefficients of incoming and outgoing currents by
\begin{equation}
\label{eigen-equation}
  \left(\begin{array}{c}
    \psi_m \\
    \psi_n 
  \end{array} \right)
  =
  \left(\begin{array}{cc}
    t^j_{mk}(E) & t^j_{ml}(E) \\
    t^j_{nk}(E) & t^j_{nl}(E) 
  \end{array}\right)
  \left(\begin{array}{c}
    \psi_k \\
    \psi_l 
  \end{array}\right),
\end{equation}
at all saddle points.
For a given potential the scattering coefficients for the saddle
points can be determined as
functions of the energy $E$ by semi-classical methods, as has been
shown and explicitly worked out by Fertig \cite{fertig}. Further, knowing
the matrices $S(E)$, the scattering conditions (\ref{eigen-equation})
form a closed system of equations for the coefficients $\psi_l$ from
which eigenenergies $E_n$ and 
eigenstate coefficients $\psi_l$ can be determined \cite{fertig}. 

Now, identifying the saddle points with nodes and the approximated
contour states  (\ref{contour-state}) with links, one
arrives at a properly defined network.
Finite isolated QH-Systems correspond to finite closed networks,
systems coupled to external leads are described by open networks.
Network states $\Psi = \{ \psi_l \}$ determine
wave functions of the real system via the expressions
(\ref{contour-state}). 
Moreover, the scattering condition (\ref{eigen-equation}) fulfilled by
eigenstates of the QH-system is just the stationarity condition of the
network; thus 
eigenenergies $E_n$ and -states $\Psi_n$ are determined by
\begin{equation}
  \label{network-equation}
  U(E_n) \Psi_n = \Psi_n,
\end{equation}
corresponding to Eq. (4.7) in Ref. \cite{fertig}.
Here the network operator $U(E)$ is energy dependent, since the
coefficients are functions of energy.

The Chalker-Coddington model is basically identical to such a network,
but exhibits some simplifications: It's nodes and links form a regular
square lattice, in contrast to saddle points and contour lines of a
real smooth disorder potential, which may give a more irregular
network. Further, for all nodes the transmission amplitudes for
the scattering into the left outgoing links is set to a constant
value $T_+$, and for the scattering into the right link to $T_-=1-T_+$.
The scattering phases $\phi_{lm}= \arg t_{lm}$ are random
variables, homogeneously distributed over $[0,2\pi]$, as in the case of the 
network model for a real disorder potential. The scattering
amplitude $T_+$ acts as model parameter.

With these simplifications the network can be viewed as one corresponding to
a QH-system with an egg-carton-like disorder potential. The basic length 
$a$ of this potential is assumed to be large compared to $l_c$.
Also it should exhibit small deviations from regularity, such 
that the flux penetrating the plaquettes fluctuates strongly on scales
of $\Phi_0$, leading to random scattering phases $\phi_{lm}$.
According to this picture we parameterize the transmission
amplitude by $T_\pm = (1 + \exp(\pm E / E_t)$, where $E$ denotes 
the electron energy (minus the cyclotron energy $\hbar \omega_c(n+1/2)$)
and $E_t$ a characteristic tunneling energy of the saddle points.
Thus, the Chalker-Coddington network is build up by nodes with
scattering matrices 
\begin{equation}
  \label{cc-matrix}
  S_j =
  \left(\begin{array}{cc}
    e^{i\fie_j(\eps)} & 0                 \\
    0                 & e^{i\fie'_j(\eps)} 
  \end{array}\right)
  \left(\begin{array}{cc}
    t_+(\eps) & t_-(\eps) \\
    -t_-(\eps) & t_+(\eps) 
  \end{array}\right),
\end{equation}
where $ t_\pm (\eps) = (1 + e^{\pm \eps})^{-1/2}$ and $ \eps = E/E_t $.
These nodes are arranged on a regular square lattice with orientation
as shown in Fig.~\ref{fig-examples}(b).
For fixed $\eps$ the
scattering phases $\fie_j(\eps),\fie'_j(\eps)$ are considered to be
random numbers 
homogeneously distributed over $[0, 2\pi]$; except for this they are not
further specified. As it has been shown by the originators, the network
exhibits delocalized states only at a singular critical point determined by
$|t_-|=|t_+|$, i.e. at $\eps_c=0$. For $\eps \neq 0$ the eigenstates
have a finite localization length $\xi$, diverging with a universal
exponent $\nu \approx 2.3$ when $\eps$ approaches $\eps_c=0$, $\xi
\propto |\eps|^{-\nu}$ \cite{chalker-coddington}.

The $\eps$-dependent network operator of the model is given by the
coefficients of the scattering matrix \Ref{cc-matrix} via 
Eq. \Ref{udef}. Then the network dynamics are
\begin{equation}
  \Psi(t) = U^t(\eps) \Psi(0), \quad  t = 0,\pm 1, \pm2, \cdots.
\end{equation}
$\eps$-dependent quasi-energies $\omega_n(\eps)$ and eigenstates
$\Phi_n(\eps)$ are defined by
\begin{equation}
  U(\eps)\Phi_n(\eps)= e^{i\omega_n(\eps)} \Phi_n(\eps),
\end{equation}
the local density of network states is
\begin{equation}
  \rho_\eps(l,\omega) = \sum_n |\phi_{n,l}(\eps)|^2
  \delta(\omega-\omega_n(\eps)).
\end{equation}

Quenched disorder is represented by the random phases $\phi_j$.
Thus the disorder average of a quantity $A$ is given by the integral
over all random-phase configurations, $[A] \equiv \int_0^{2\pi} 
\prod_j\frac{d\fie_j }{ 2\pi}  A\{\fie_m\} $. As usual, in numerical
calculations this ensemble average is approximated by the average over
a finite set of disorder configurations $\{\fie_m\}$. In case of local
observables $A_l$ the statistics are often improved by taking also a
system average $[A]_S = N^{-1} \sum_{n=1}^N A_n$.

A generalized version of this model is obtained, when also the
transmission amplitudes become random variables, e.g. by
$t^j_\pm = ( 1 + \exp(\eps - \mu_j))^{-1/2}$, where $\mu_j \in
[-W/2,W/2]$ is random and independent for each node $j$. This type
of disorder corresponds to fluctuating saddle point energies 
of the random potential, which brings in aspects of classical 
percolation \cite{lee,klesse-metzler,Metzler1}.

\begin{figure}[htb] 
\begin{center}
        \epsfxsize 8cm
        \epsffile{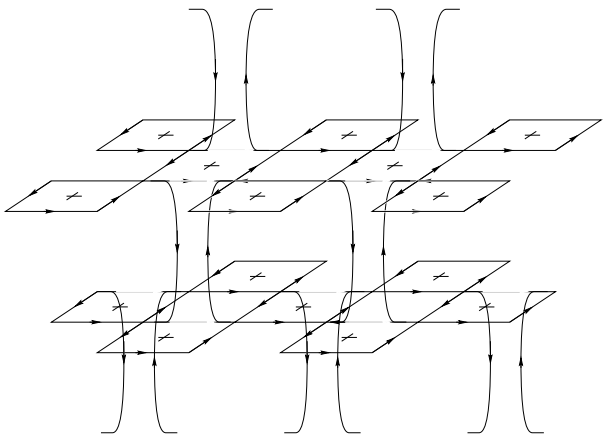}
\end{center}
\caption{ Structure of the 3D network model.  }
\label{fig-3dnetwork}
\end{figure}

The 3D model we will investigate is very similar to that of Chalker and 
Dohmen \cite{chalker-dohmen}. It consists of a stack of
the 2D Chalker-Coddington networks we just have  considered, now
coupled into the 
vertical direction by interlayer links as shown in
Fig.~\ref{fig-3dnetwork}.
However, the local scattering matrices are defined as in the
2D network; so here 
the scattering amplitudes are also determined by the energy-parameter $\eps$
and the disorder is represented by random scattering phases.
Due to the inter-layer coupling, the singular
delocalized point at $E=0$ of the 2D network becomes an extended
metallic band ranging from $-E_c$ to $E_c$, $E_c >0$. Outside this
band the states are localized \cite{chalker-dohmen}.

\section{Phase coherent Diffusion in Disordered Systems}
\label{sec-dynamics}
The standard method to determine numerically a quantum particle's time
evolution is first to calculate all eigenvalues $E_n$ and eigenvectors
$\phi_n$  of the Hamiltonian. 
After that, the state at any time $t$ of a particle
initially in state $\psi(0)$ can be calculated simply by
\begin{equation}
  \psi(t) = \sum_n e^{-iE_nt} \langle \phi_n, \psi(0) \rangle.
\end{equation}
In this procedure, the diagonalization of the Hamiltonian takes the
main part of the numerical effort and limits the application of this
method to rather small system sizes
\footnote{For example, the Hamiltonian for a 3D system of $L\times L
  \times L =50 \times 50 \times 50$ lattice points possesses $50^3$
  energy levels and as many eigenvectors with $50^3$ complex entries
  each. This yields approximately $(50^3)^2 \approx 1.6 \times 10^{10}$
  complex numbers, equivalent to at least about 100 Giga Byte, which
  to calculate or to back up is a hard job, also for todays
  computers. 
}.

The dynamical network model
offers a far better way to simulate quantum mechanical time evolution.
Using the network dynamics Eq.(\ref{dynamic}), one
can evolve
any arbitrary state $\Psi(0)$ in discrete time steps by
iterative multiplication of $\Psi(0)$ with the network operator $U$, 
\begin{equation} \label{formula}
  \Psi(t) = U^t \Psi(0) = U \:\Psi(t-1), \quad t=0,1,2, \cdots, 
\end{equation}
Since multiplication of
vectors by matrices is one of the simplest tasks computers can perform
(particularly when most of the matrix entries vanish, as in our case)
the dynamics are very suitable for numerical applications. In fact, by
this the simulation of the quantum mechanical time evolution 
becomes as easy as the simulation of a classical system with discrete
dynamics.

Let us consider numerical calculations for the 2D and 3D networks
described in the last section. 
In both cases we investigate the spreading of network states 
initially confined to a few links in the center of a squared (cubic)
network of volume $L^d$.
The diffusion of such states is described by the disorder averaged
($[\dots]$) probability density
\begin{equation}
 p(r,t) = \left[ |\psi(r,t)|^2 \right].
\end{equation}
($\psi(r,t)$  understood as the link amplitude $\psi_l(t)$ at link $l$
closest to coordinate $r$.)
We will determine this density for states at energies in the
metallic, the insulating and in the transition regime. The latter
regime is especially interesting, since at this energy interference
effects start to destroy the diffusion \cite{anderson}.

Before the setups are explained in detail and results are
presented, let us briefly summarize some results concerning quantum
diffusion in disordered systems. For the transition point, these
results are gained mainly from scaling theory 
\cite{chalker-daniell,brandes}.

At energies deep in the metallic regime
the density $p(r,t)$ is given by the Gaussian density,
\begin{equation}
  \label{gaussian}
  p(r,t) = \frac{1 }{ (4 \pi D t)^{d/2}} e^{-r^2/4Dt},
\end{equation}
where $D$ is the diffusion constant (ordinary diffusion). Accordingly,
the $q$th moment of the density 
grows like
\begin{equation}
  M_q(t) \equiv \int d^dr \:r^q\:p(r,t) \propto t^{hq}
\end{equation}
with exponent $h=1/2$.

In the vicinity of the transition point, scaling theory 
predicts an algebraic decay of the density at short distances
$r \ll  (t/\rho_c)^{1/d}$
\begin{equation}\label{rho_critical_b}
  p(r,t) \propto t^{-D_2/d} r^{D_2-d},
\end{equation}
where $D_2$ is the correlation dimension of critical
eigenstates and $\rho_c$ the density of states at the transition point
\cite{abrahams,chalker-daniell,brandes}.  
At large distances,  $r \gg  (t/\rho_c )^{1/d}$, the decrease of the
density is still exponential. 
>From this follows that at the critical point the $q$th moment scales
like  \cite{brandes}
\begin{equation}
 M_q(t)  \propto t^{q/d}.
\end{equation}
Therefore, in $d > 2$ dimensions,  the exponent $h$ shrinks down from
$1/2$ in the metallic regime to the lower critical value $1/d$ when the
transition point is 
approached, indicating the transition to the localized regime, where
diffusion ceases. 

In the localized regime, the wave-packet spreads out until its size has
become of the order of the localization length $\xi$. Up to this point
the density $p(r,t)$ evolves in the same way as at the critical
point. At larger times, $t \gg \rho_c \xi^d$, the density
equilibrates in a volume $\xi^d$, centered around the starting point.

Now, let us consider the outcomes of our computer simulations. We
begin with the 
diffusion on the Chalker-Coddington network at critical energy $E_c=0$.
The initial state $\Psi(0)$ is located
on four links labeled by 0,1,2,3 in the center of a squared closed
network with periodic boundary conditions,
\begin{equation}
  \Psi(0) = \frac{1}{ 2}( {\bf e}_0+{\bf e}_1+{\bf e}_2+{\bf e}_3).
\end{equation}
This state is subjected to the time evolution Eq.~(\ref{dynamic}),
whereby $E=E_c$.
Figure \ref{fig-sequence} visualizes the diffusion of a particle
on a $40\times 40$ node network and 
also demonstrates the numerical precision of the simulations.
\begin{figure}[htb] 
  \begin{center}
        \epsfxsize 8cm
  \end{center}
  \caption{A state evolving backwards after it has moved 350 time
    steps  forward.
    The initial state is reconstructed to a high accuracy (last picture of
    the lower row).
    }
  \label{fig-sequence}
\end{figure}
The first picture in the upper row shows the density $|\Psi(r,t)|^2$
after 350 time steps. The initial state $\Psi(0)$
has been almost equilibrated over the whole lattice. In
order to test that this equilibration is by no means influenced or
even caused by accumulating numerical errors, we applied successively
the adjoint operator $U^+=U^{-1}$ on the final state
$\Psi(350)$. Doing this, the state evolves back (see
Fig.~\ref{fig-sequence}) and returns after another 350 steps to the initial
state with high accuracy, $|\Psi'-\Psi| < 10^{-14}$.
\footnote{The stability of the quantum mechanical 
time evolution against numerical errors seems to be 
quite astonishing, since it is
in sharp contrast to the numerical instability of the classical
counterpart. 
Because of the chaotic dynamics, in the latter the
numerical uncertainty in position and impulse makes the
trajectory of a classical particle completely indefinite after a few
time steps.
However, the dynamical stability of quantum-chaotic 
systems is well-known and has been first observed by Shepelyansky
[30].}

\begin{figure}[htb] 
\begin{center}
        \epsfxsize 9cm
        \epsffile{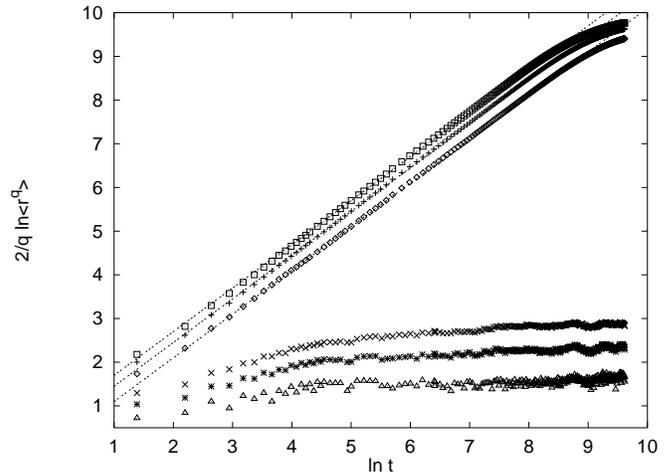}
\end{center}
\caption{The 2nd ($\Diamond$), 4th ($+$) and 6th ($\Box$) moment (to
        the power of $2/2$ 
  $2/4$ and $2/6$, respectively) of
  spreading wave-packets at the transition point at energy
 $E=E_c=0$ and in the strong
  localized regime, $E=3E_t$ (where $\triangle$, $\ast$, $\times$ 
correspond to the 2nd, 4th and 6th moment). The dashed lines have slope 1.}
\label{fig-moments}
\end{figure}

Fig.~\ref{fig-moments} shows the $q$th moment $M_q(t)$ to the power
of $2/q$ for $q=$2,4 and 6 at critical energy $E_c=0$ and in 
the strong localized regime at energy $E= 3E_t$.
The data for $E_c$ are gained from 5 networks of 300$\times$300 nodes with
periodic boundary condition. Since the moments are obtained by
integration over the entire system, their fluctuations with
respect to different disorder configurations are quite small;
the average of 5 configurations yields moments
proportional to $M_q(t)\propto t^{qh}$, with an exponent $h$ very close
to $1/d=1/2$ (see Fig. \ref{fig-moments}).
At times greater than the diffusion time of order $\ln t_L =
\ln(L/2)^2 \approx 9$ the increase goes down, since then the extension
of the wave-packet has reached the system size $L$.
Away from the critical point, the moments saturate quickly at certain
values, reflecting the absence of diffusion at length scales greater
than the localization length.

\begin{figure}[htb] 
\begin{center}
       \epsfxsize 9cm
       \epsffile{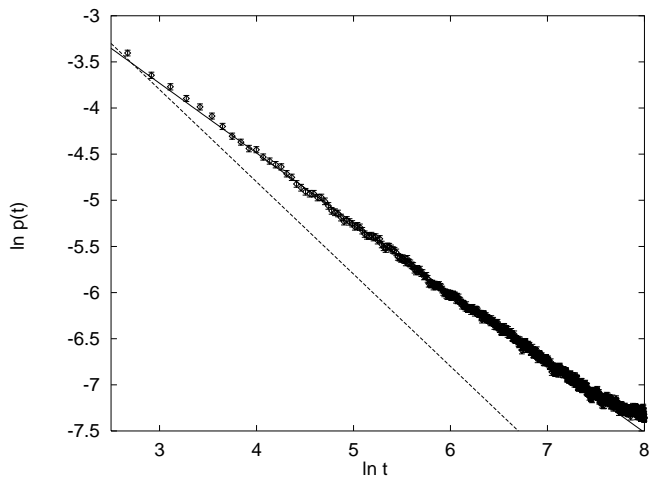}
\end{center}
\caption{The averaged return probability $p(t)$ taken from 400
  networks of $100\times 100$ nodes each. The data fits to a straight
  line with slope $x = 0.76\pm 0.01$. The dashed straight line with
  slope -1 corresponds to normal diffusion.}
\label{fig-pvont_2d}
\end{figure}

Next, we consider the probability of the particle to return
to the initial volume formed by the links $l=0,1,2,3$,
\begin{equation}\label{return_prob}
  p(t) =  \sum_{l=0}^3 |\psi_l|^2.
\end{equation}
Since this return probability fluctuates strongly with respect to
disorder, it is
necessary to average over many configurations.
The curve plotted in Fig.~\ref{fig-pvont_2d} results from 400 different
critical networks of $100\times 100$ nodes each. The return
probability decays with a power law, $p(t) \propto t^x$, with
exponent $x = 0.76 \pm 0.01$. 

In accordance to Eq. (\ref{rho_critical_b}), this exponent fits 
well to the correlation dimension for critical states, which is
exactly twice that,
$D_2/2 = 0.75 \pm 0.025$ (see
Sec. \ref{sec-statics}). The straight line with slope -1 corresponds to
the decay of  the return probability one would expect for classical 
diffusion, governed by Eq. (\ref{gaussian}).

\begin{figure}[htb] 
\begin{center}
         \epsfxsize 8cm
         \epsffile{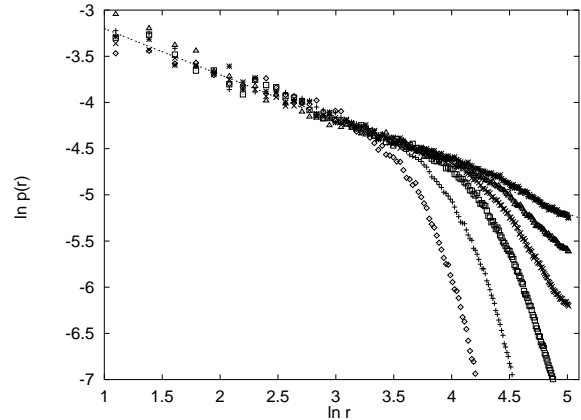}
\end{center}
\caption{The rescaled density $t^{D_2/2}\bar p(r,t)$ for times
  $t=1500$ ($\Diamond$) up to 15000 ($\ast$). The straight lines
  corresponds to $r^{D_2-d} 
  \approx r^{-0.5} $.}
\label{fig-radial}
\end{figure}

Finally, we discuss the density $p(r,t)$ for fixed times $t$ as a
function of $r$. In
order to reduce the numerical effort, we average $p(r,t)$ over a
ring of radius $r$ around the origin:
\begin{equation}
  \bar p_t(r) = \frac{1}{2 \pi} \int p(|r|,\varphi,t) d\varphi.
\end{equation}
Taking for granted that the disorder averaged density
$p(|r|,\varphi,t)$ is radially symmetric, the new density $\bar p_t(r)$
exhibits the same scaling behavior as $p(r,t)$. However, due to
the integration over the ring $\bar p_t(r)$ fluctuates much less
than $p(r,t)$, which considerably reduces the number of
configurations necessary for a satisfyingly small statistical error. Because of
Eq. (\ref{rho_critical_b}), the curves $\bar p_t(r)$ for different
times $t$ should coincide for small radii $r$ after rescaling them by
\begin{equation}
  \bar p_t(r) \longrightarrow t^{D_2/2} \bar p_t(r).
\end{equation}
The averaged density $\bar p_t(r)$ of five critical systems
of 300$\times$300 nodes and times $t$ ranging from $t=1500$ up to $15000$
steps shows that this is indeed the case (Fig.~\ref{fig-radial}).
The algebraic decay of the density for small $r$
changes to an exponential decay at radii larger than a crossover
length, which increases with time $t$. The data plotted
in Fig.~\ref{fig-radial} for small $r$ fits to $r^{D_2-d}$,
with $d-D_2=0.5$, as predicted by Eq. (\ref{rho_critical_b}).

\begin{figure}[htb] 
\begin{center}
         \epsfxsize 8cm
         \epsffile{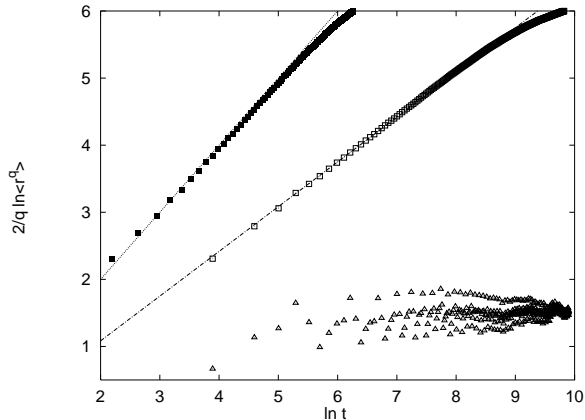}
\end{center}
\caption{ The second moment $M_2(t)$ at energies
  $E=1.5$ (filled $\Box$), 3.4 ($\Box$) and 6 ($\triangle$), 
  corresponding to the metallic, transition and isolating regime of
  the 3D network model, respectively.}
\label{fig-momente_3d}
\end{figure}

For the 3D model we carried out analogous simulations.
We first determined the moments $M_q(t)$ for states initially
restricted to a plaquette in the center of a cubic network of
$L^3=50^3$ nodes. In Fig.~\ref{fig-momente_3d} the moments
$M_q(t)^{2/q}$ are plotted for $q=2,4,6$ at energies $E=1.5, 3.4$ and
6, measured in units of the tunnel energy $E_t$. 
The average of $5$ disorder configurations has been taken. At energy
$E=1.5$ the moments scale with an exponent $h=1/2$, indicating
pure metallic diffusion according to Eq. (\ref{gaussian}), for $E=6$
the moments converge to finite values small compared to $L^q$, which
shows clearly the absence of diffusion, i.e. localization.
For energies $E = 3.4 \pm 0.1$ we found an anomalous
reduced exponent $h$ very close to $1/3$. Consequently, energy
$E=1.5$ must be in the metallic and $E=6$  in the strongly localized
regime. The transition energy we determined to be $E_c=3.4\pm 0.1$,
since at energies 3.3 and 3.5  significant deviations from the
critical scaling $M_q(t)\propto t^{q/3}$ are already visible.

\begin{figure}[htb] 
\begin{center}
       \epsfxsize 8cm
       \epsffile{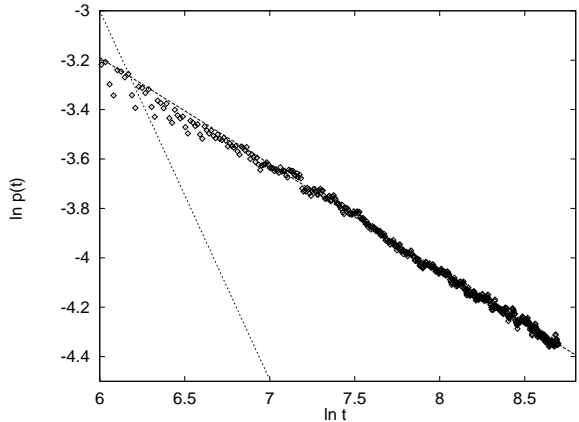}
\end{center}
\caption{ The critical return probability decays proportional $\propto
  t^{-D_2/3}$, $D_2/3 = 0.43 \pm 0.04$. Data obtained from 200 systems
  of $50^3$ nodes each. The dotted line with slope $-3/2$ corresponds
  to normal diffusion.}
\label{fig-pvont_3d}
\end{figure}

The return probability $p(t) \propto r^{-x}$ shown in Fig.~\ref{fig-pvont_3d}
results from 200 systems of size $L^3=50^3$ at critical energy
$E=3.4$.
 From these data we obtained $x=0.43 \pm 0.04$ and thus
$D_2 = 3x = 1.3 \pm 0.12$  according to Eq. (\ref{rho_critical_b}).
For the 3d Anderson transition this correlation dimension 
has not yet been determined very precisely. The values 
obtained in previous publications are $D_2 = 1.7 \pm 0.3$
\cite{soukoulis}, $ D_2 \approx 1.45 - 1.8 $ \cite{schreiber} and
$D_2 = 1.7 \pm 0.2$ \cite{brandes}. An analytical calculation using an epsilon
expansion yields $D_2 = 2 - \epsilon = 2 - ( 3 - 2 ) = 1$ \cite{wegner}.
It should be noted that the network model corresponds to a system with
a magnetic field while in Ref. \cite{soukoulis,schreiber,brandes}
time reversal systems in the absence of magnetic fields
were considered. 

\begin{figure}[htb] 
\begin{center}
    \epsfxsize 8cm
    \epsffile{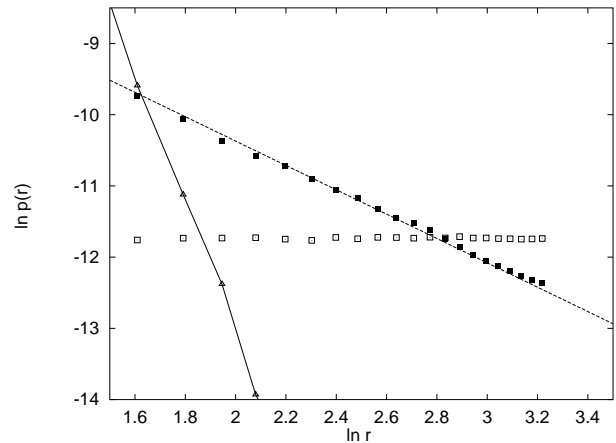}
\end{center}
\caption{The density $\bar p(r,t)$ at time $t=20000$ of states initially
  located in the center of a $3D$ network of $50^3$ nodes.
  The curves correspond to energies $E=1E_t$ (metallic, $\Box$),
  $E_c=3.4E_t$ (critical, filled $\Box$) and $E=6E_t$ (localized
  regime, $\triangle$).
}
\label{fig-radial_3d}
\end{figure}

Finally, Fig.~\ref{fig-radial_3d} shows the density profile $\bar
p(r,t)$ at time $t=20000$ of states initially located in the center of
the network in the metallic, critical and localized regime. At this
time, metallic states at $E=1$  have been completely
equilibrated over the entire system, seen by the constant density
$\bar\rho(r,t)$, whereas the density of states at $E=6$ in the
localized regime shows a strong, exponential decay.
Both densities do not evolve further in time. At the transition point,
just at time $t=20000$ the bulk of the spreading wave-packets cover the
whole lattice. Therefore, the corresponding density decays
algebraically $\propto r^{-y}$ according to
Eq. (\ref{rho_critical_b}) with exponent $y$ close to $\eta =
3-D_2 = 1.7 \pm 0.1$.

To summarize the outcome of the simulations, both the 2D and 3D
dynamical network models show precisely the expected diffusive
behavior. The metallic regime of the 3D model shows normal diffusion
with a Gaussian density $p(r,t)$. At the critical point in both
systems the simulations fit very well to the anomalous
diffusion predicted by scaling theory.
It is also quite remarkable that the simple, locally defined
dynamics (\ref{dynamic}) lead immediately to the well-known but highly
non-trivial Anderson-localization, caused by long ranged destructive
interferences. 

Since dynamical network models correctly describe the dynamics of
disordered systems in the metallic, localized and even in the
transition regime, their generating operator $U$ must necessarily
contain relevant information on the spectral properties in the
corresponding regimes. Certainly, this information is encoded in the
quasi-spectrum $\{ \omega_n \}$ of $U(E)$, on which we will focus now.

\section{Spectral Statistics}
\label{sec-spectral}
We start with the investigation of the {\em local} spectral density
$\rho_0(\omega)$ of the two models under consideration. After that we
turn over to the statistics of quasi-energy-level
distribution in the Chalker-Coddington network. As before, particular
attention is paid to the metal-insulator transition point.
 
\subsection{Local Spectral Density}
\label{subsec-ldos}
The local spectral density can be calculated efficiently by iteration
in nearly the 
same way as before the return probability $p(t)$.
Instead of the return probability Eq. \Ref{return_prob} one has to
determine the complex amplitude $\psi(0,t) = \langle {\bf e}_0, U^t
{\bf e}_0 \rangle $, i.e. the propagator $G_{00}(t)$. In principle,
Fourier-transforming the latter immediately yields the spectral
density $\rho_0(\omega)=\sum_n |\Phi_n(0)|^2\delta(\omega-\omega_n)$
because of
\begin{eqnarray}
  G_{00}(t) = \langle \bfe_0, U^t \bfe_0 \rangle &=& \sum_n
  e^{i\omega_nt} |\Phi_n(0)|^2 \\
  &=& \int d\omega\:e^{i\omega t}   \rho_0(\omega).
\end{eqnarray}
Thus, to obtain the local spectral density $\rho_0(\omega)$ we
generate the sequence $G_{00}(t), t=0,1,\dots T$ successively by
applying $U$ on the state $\bfe_0$. Before doing the numerical Fourier
transformation, we softened the sharp edges of the time-window from 0
to T by multiplication with $1/2-1/2\cos(2\pi t/T)$ (the so called
Hamming window), in order to reduce artifacts due to the finite time
range. Additionally, to ensure numerical stability, we damped the
Green function with the exponential $e^{-t/\tau}$, whereby the damping
time $\tau$ fulfills $1 \ll \tau \ll T$.  Then, performing a discrete
Fourier transformation on this windowed and damped Green function and
taking the real part of the result we get the quantity
\begin{equation}
  \rho'_0(\omega)= \sum_n |\phi_n(0)|^2
  \delta_\tau(\omega-\omega_n), \quad \omega_n = \frac{2\pi }{T} n,
\end{equation}
where $\delta_\tau(\omega)$ is a peaked function with maximum value at
$\omega = 0$ and width of order $1/\tau$. For sufficiently large times
$\tau$ and $T \gg \tau$, this quantity converges to the wanted density
$\rho_0(\omega)$.

\begin{figure}[htb] 
\begin{center}
       \epsfxsize 8cm
       \epsffile{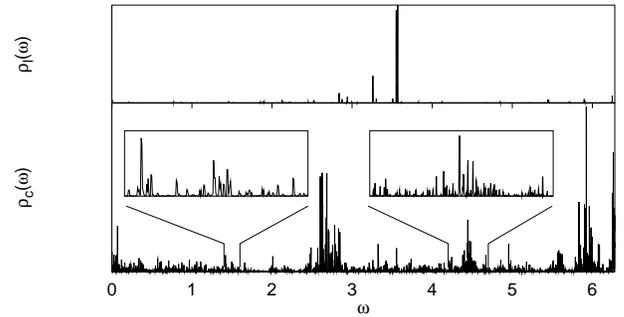}
\end{center}
\caption{Local spectral density $\rho_0(\omega)$ of the 2D
  Chalker-Coddington network of $50\times 50$ nodes in the localized
  regime at $E=2E_t$ (upper) and at the transition point (lower)
  at $E=E_c=0$. The localized density is concentrated on few
  frequencies, indicating strong localization. The critical density is
  distributed over the entire band and appears to be scale invariant.}
\label{fig-spectrum_2d}
\end{figure}

By this method, the spectral densities plotted in
Fig.~\ref{fig-spectrum_2d} of a Chalker-Coddington network of
$50\times 50$ nodes have been calculated, whereby we chose $\tau =
2000$ and $T=32768$.  In the upper spectrum, at energy $E=2E_t$, the
density is almost concentrated at only a few energies, reflecting strong
localization. The lower, critical spectrum at $E_c=0$ fluctuates
strongly and turns out to be scale invariant, as we will see in a
moment.

\begin{figure}[htb] 
\begin{center}
       \epsfxsize 8cm
       \epsffile{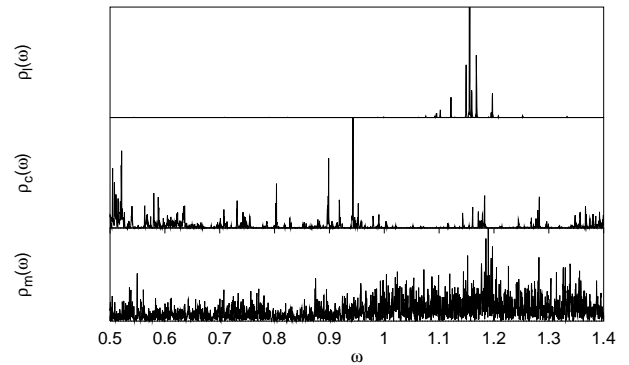}
\end{center}
\caption{Local spectral densities of a 3D network model of $50^3$
  nodes in the localized ($E=6E_t$), critical ($E=3.4$) and
  metallic ($E=1$) regime.}
\label{fig-spectrum_3d}
\end{figure}

The same method applied to the 3D network model considered
above gave the three spectra shown in Fig.~\ref{fig-spectrum_3d},
corresponding to the localized, critical and delocalized regime.
For the calculations we used systems of $50\times 50\times 50$ nodes,
a maximum time $T=32768$ and $\tau=4000$. At energy $E=6$ in the
localized regime and at the transition point $E_c=3.4$ the spectra are
qualitatively similar to the corresponding ones of the 2D model, whereas at
energy $E=1$, where states are delocalized, the spectrum is
comparatively homogeneous.

\begin{figure}[htb] 
\begin{center}
    \epsfxsize 8cm
    \epsffile{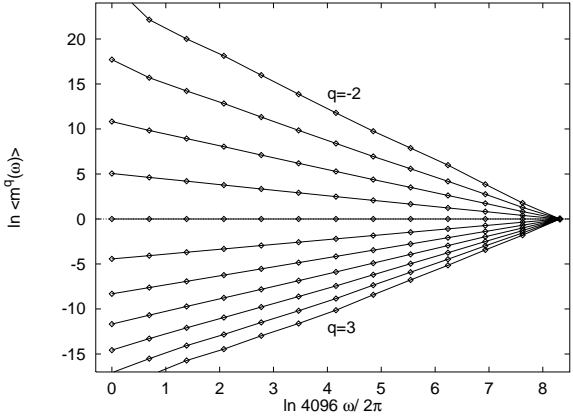}
\end{center}
\caption{Scaling behavior of $q$th powers of masses $m(\Omega)=
  \int_{\omega_0}^{\omega_0+\Omega} \rho_0(\omega) \:d\omega $ for box
  sizes $\Omega$ from $2\pi/4096$ up to $2\pi$ and powers $q$ ranging
  from -2 to 3 in steps 0.5. The data result from the average over non
  overlapping intervals of 28 systems of $200\times200$
  nodes each.}
\label{fig-boxscaling}
\end{figure}

\begin{figure}[htb] 
\begin{center}
    \epsfxsize 8cm
    \epsffile{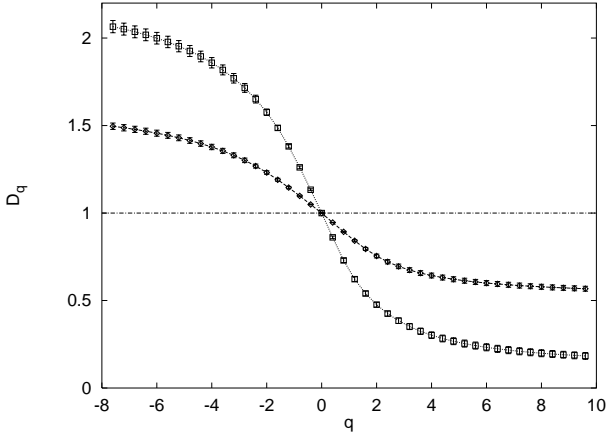}
\end{center}
\caption{Generalized dimensions of critical local spectral densities
  in the 2D ($\Diamond$) and 3D ($\Box$) network model.}
\label{fig-spectrum_mfa}
\end{figure}

The critical spectral densities deserve a closer look. Their scale
invariance (or self-similarity) can be demonstrated by examining the
scaling behavior of the quantity $m(\Omega) =
\int_{\omega_0}^{\omega_0+\Omega} \rho_0(\omega) \:d\omega$, which is
the mass of the interval $[\omega_0, \omega_0+\Omega]$ with
respect to the density $\rho_0(\omega)$. Disorder averaged powers of
it, $m_q(\Omega)= \langle m^q(\Omega) \rangle$, scale over a wide
frequency ($\Omega$) range with definite exponents, as can be seen by
the linear dependence of $\ln m_q(\Omega)$ on $\ln \Omega$, plotted in
Fig.~\ref{fig-boxscaling}.  This power law scaling of $m^q(\Omega)$
proofs the scale invariance of the density $\rho_0(\Omega)$.
Unlike in the case of homogeneous densities, the scaling exponent depends
nonlinearly on $q$, therefore the densities are called multifractals
\cite{janssen}.  They can be characterized by generalized dimensions
$\tilde D_q$
describing the scaling of the $m_q(\Omega)$ by $m_q(\Omega) \propto
\Omega^{d+(q-1)\tilde D_q }$, for, in principle, all real $q$. $d$ denotes
the dimension of the density's support, thus $d=1$ for the spectral
measure.  \footnote{ Equivalently, instead of the generalized
  dimension $\tilde D_q$ often also exponents $\tau_q \equiv (q-1)\tilde D_q$ or the
  Legendre transformed $f(\alpha)$ of the function $\tau_q$ are used
  to describe a multifractal density \cite{janssen}.} Figure
\ref{fig-spectrum_mfa} shows the generalized dimensions of the
critical spectral densities of both, the 2d and the 3d model. For the
determination of the $\tilde D_q$ we use 28 2d systems of $200\times 200$
nodes and 5 3d systems of $50\times 50 \times 50$ nodes, respectively.
The values for some characteristic exponents, namely the information
dimension $\tilde D_1$, the correlation dimension $D_2$, $D_{-\infty}$,
$D_\infty$, and the quantity $\tilde\alpha_0 = d - \partial\tilde D_q/
\partial q(q=0)$ are 
listed in Tab.~\ref{tab-mfa}.
\begin{table}[t]
  \begin{tabular}{c|c|c|c|c|c} 
    $\mbox{Dim.}$&$ \tilde D_1$&$ \tilde D_2$&$ \tilde D_{-\infty} $&$
    \tilde D_\infty $&$\tilde \alpha_0$\\ 
    \hline 
    $2d $&$ 0.87\pm0.01 $&$ 0.75\pm0.01 $&$1.5\pm0.2 $&$ 0.6 \pm 0.1$&$
    1.14 \pm 0.01$\\ 
    \hline 
    $3d $&$ 0.67\pm 0.02 $&$ 0.48 \pm 0.03 $&$ 2.1 \pm 0.1 $&$
    0.2 \pm 0.2 $&$ 1.34 \pm 0.2 $\\ 
  \end{tabular}
  \caption{Generalized dimensions and $\tilde \alpha_0$ of critical local
    spectral densities in the 2D and 3D network model.\label{tab-mfa} }
\end{table}
The coincidence of the correlation dimension $\tilde D_2 = 0.75 \pm 0.02 $
and $\tilde 0.48 \pm 0.03$ respectively with the decay exponents of the
return probability $p(t)$, $x=0.76 \pm 0.01$ and $0.43 \pm 0.04$ is not by
accident, but 
because $\rho_0(\omega)$ and $p(t)=|\psi_0(t)|^2$ are related via
$\psi_0(t)$ by Fourier transformation, which implies that $x$
equals $\tilde D_2$ \cite{ketzmerick}.

\subsection{Quasi-Energy Level Distribution in the Chalker-Coddington
Network} 
\label{subsec-level}
By interpreting $U$ as a time evolution operator,
the eigenphases $\omega_l$ of the unimodular eigenvalues
$e^{i\omega_l}$ of $U$ must be considered as the analog to energy levels.
In the following we will consider the distribution of these eigenphases
or quasi-energy levels $\omega_l(\eps)$ of the Chalker-Coddington network.
We will determine the level spacing distribution
function $P(s)$ and the level number variance $\Sigma_2(N)$. $P(s)ds$
gives the probability to find a quasi-energy level at distance $sds$
of another level, whereby $s$ is measured in units of the average
level spacing $\Delta$. $\Sigma_2(N)\equiv \langle (n - \langle n
\rangle )^2 \rangle $ denotes the variance of the 
number of levels in an interval which contains on average $\langle n
\rangle = N$ levels.

The investigated systems are closed networks with periodic boundaries
of $L\times L=50\times 50$ nodes in the localized and critical regime
at energies  $E=10$ and $E=E_c=0$, respectively.
The corresponding unitary network
operators $U(E)$ are represented by $M=2\times50\times50$
dimensional matrices.
Diagonalizing such a matrix by standard numerical
methods yields $M$ complex eigenvalues $e^{i\omega_l(E)}$,
homogeneously distributed on the unit circle. The corresponding
quasi-energy levels $\omega_n(E)$ are homogeneously distributed in
the interval $[0,2\pi[$ with an averaged spacing $\Delta = 2\pi/M$. It
is not necessary to perform special unfolding-procedures in order to
compensate inhomogeneities in the distribution. 
For these quasi-level distributions $\omega_l(E)$ we determine
$P(s)$ and $\Sigma_2(s)$.

\begin{figure}[htb] 
\begin{center}
    \epsfxsize 9cm
    \epsffile{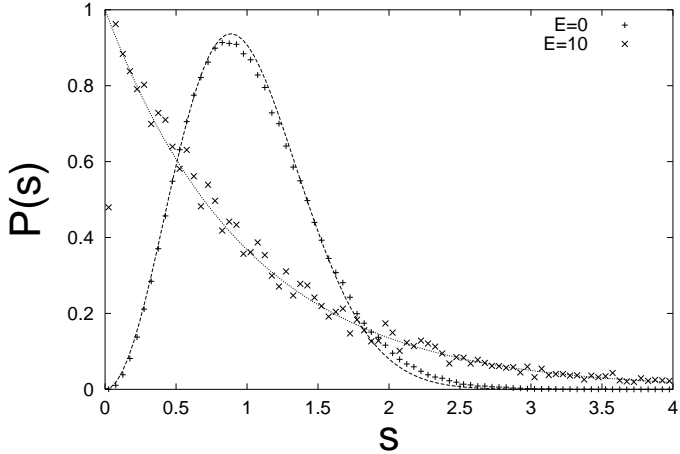}
\end{center}
\caption{The level distribution function $P(s)$ 
  for quasi-levels of   networks in the localized regime, $E=10$
  ($\times$), 
  and at the transition point $E=E_c=0$ ($+$).
  In the localized regime the distribution is clearly governed by
  Poisson statistics (solid curve), whereas it is close to the
  Wigner surmise for a unitary ensemble (dashed line) at the
  transition point. 
 }
\label{fig-pvons}
\end{figure}

\begin{figure}[htb] 
\begin{center}
    \epsfxsize 9cm
    \epsffile{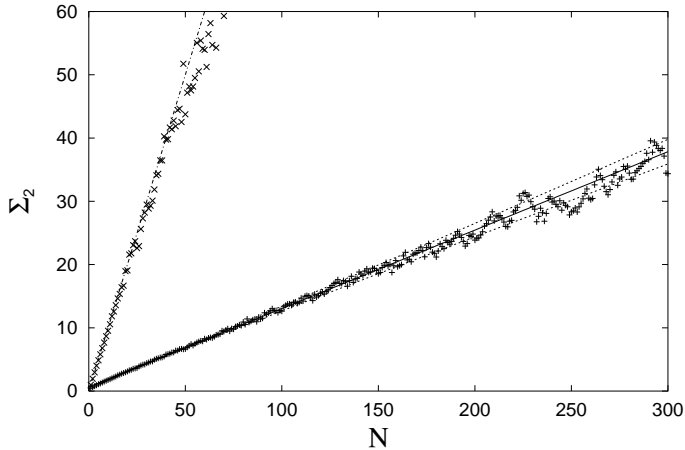}
\end{center}
\caption{
  Level number variance $\Sigma_2(N)$ for
  quasi-levels of   networks in the localized regime, $E=10$
  ($\times$), and at the transition point, $E=0$ ($+$).
 }
\label{fig-sigma}
\end{figure}

Fig.~\ref{fig-pvons} shows the level distribution function $P(s)$
for networks in the localized regime at energy $E = 10$ and
at the transition point at $E=E_c = 0$.
Since for $E = 10$ the networks states are strongly localized,
their eigenvalues (quasi-energies) are uncorrelated, reflected by the
Poisson distribution $P(s) = \exp(-s)$. Accordingly, the level
number variance is given by $\Sigma_2(N) = N$ (Fig. \ref{fig-sigma}). 

At the transition point, eigenstates are extended giving rise to
a strong level repulsion, as can be seen by the corresponding
level distribution function, which resembles the Wigner surmise for a
unitary ensemble (Fig. \ref{fig-pvons}). 
However, Fig \ref{fig-sigma} shows that the level number
variance is still linear in $N$, $\Sigma_2(N) = (0.124\pm0.006)\times N$.
This is in contrast to metallic behavior, where $\Sigma_2(N) \propto
\ln(N)$, but resembles the localized phase where $\Sigma_2(N) = N$.
The difference to the latter is only the reduced proportionality
constant, the spectral compressibility, $\chi = 0.124 < 1$. This
value of the spectral compressibility fits to the result of 
Chalker et al. \cite{chalker-lerner-smith}, according to which $\chi =
(2-D_2)/2d$, $D_2$ being 
the correlation dimension of critical states in Quantum Hall systems
\cite{klesse-metzler2}. 
The non-vanishing compressibility is reflected by a longer tail in the
level spacing distribution. As predicted by Altshuler et al. \cite{Altshuler} 
the tail of $P(s)$ decreases as $\exp(-\kappa s)$, where 
$\kappa=1/(2\chi)\approx 4$.
In \cite{Batsch_Schweitzer,metzler_varga,Metzler2} this relation could be verified
for the Anderson model and the Chalker-Coddington model, respectively.  
Since there is no real metallic phase in the QHE, 
we cannot give examples for metallic quasi-level distributions from
the Chalker-Coddington network.

\begin{figure}[htb] 
\begin{center}
    \epsfxsize 9cm
    \epsffile{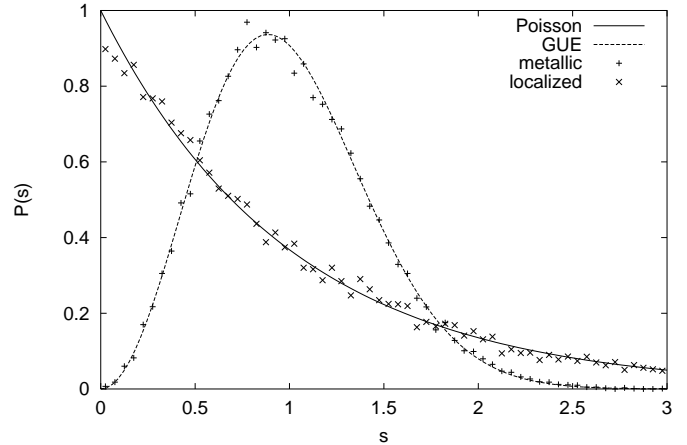}
\end{center}
\caption{
  Level distribution function $P(s)$ 
  for quasi-levels of 3d networks in the localized ($E=10$, $\times$ )
  and metallic ($E=0$,$+$) regime.
  The data fit well to Poisson statistics (solid curve)
  and to the Wigner surmise for a unitary ensemble (dashed line),
  respectively.
 }
\label{fig-pvons3d}
\end{figure}

\begin{figure}[htb] 
\begin{center}
    \epsfxsize 9cm
    \epsffile{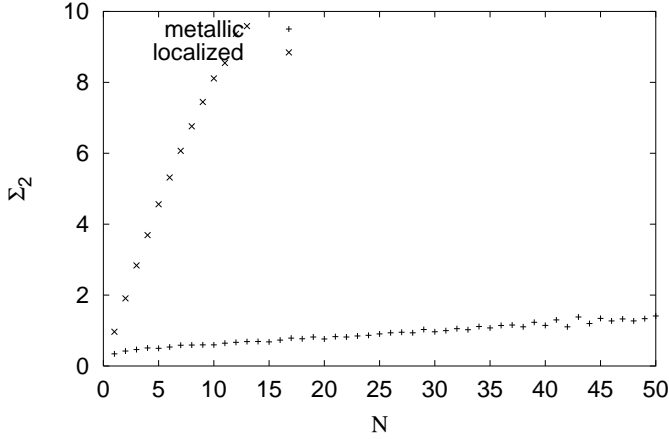}
\end{center}
\caption{
  Level number variance $\Sigma_2(N)$ for
  quasi-levels of 3d networks in the localized ($\times$) and the
  metallic ($+$) regime.
 }
\label{fig-sigma3d}
\end{figure}

For the 3D network this is possible. Fig \ref{fig-pvons3d} shows
that the levels spacing distribution in the localized and
metallic regime at energies $E=0$ and $E=10$ agree well
with Poisson statistics and the Wigner surmise for a unitary ensemble,
respectively. (Data obtained from
40000 quasi-energy levels of $10\times 10 \times 10$ nodes networks.)
The corresponding level number variances plotted in
Fig. \ref{fig-sigma3d} indicates Poisson statistics in the
localized regime and show the strongly suppressed variances for the 
metallic regime.

These two examples demonstrate that the distribution 
of network specific quasi-energy levels is
equivalent to the distribution of real energy levels,
as could be anticipated from the fact that network models
correctly reproduce the dynamical properties of real physical systems.

Another argument supporting this equivalence relies on the symmetries
and homogeneity of the flow of the quasi-energy levels $\omega_l(E)$
as a function of the real energy $E$. It can be found in 
\cite{klesse-metzler2}.

\section{Eigenstates of the Network} 
\label{sec-statics}
Correlations between eigenvalues of random matrices come along with
certain correlations in the amplitudes of eigenstates,
which will be the subject of this section. As before, we investigate
correlations of the Chalker-Coddington network, whereby we confine 
ourself to the critical region. Our findings for the network model
will be compared to results obtained within other, conventional
models, describing Quantum Hall systems with short range disorder
by a tight-binding Hamiltonian \cite{tight-binding} and  
by a so-called random Landau model \cite{random-landau}.

Network eigenstates  $\Psi_n$ at energy $E$ are solutions of the 
eigenvalue problem
\begin{equation}
  \label{eigenvalue}
  U(E)\Psi_n=e^{i\omega_n}\Psi_n.
\end{equation}
Notice that the energy $E$ is fixed and only the quasi-energy
$\omega_n$ varies. A particular eigenstate $\Psi_n$ with link
amplitudes  $\{\psi_l \}_{l=1,\cdots N}$ refers to a spatial
wave function $\Psi(r)$ via the link locations $r_l$ by identifying
$\Psi(r_l) \equiv \psi_l$. This rather vague definition is 
satisfactory  as long as the 
considered length scales are large compared to microscopic scales.

\begin{figure}[t]
  \begin{center}
    \leavevmode
    \epsfig{figure=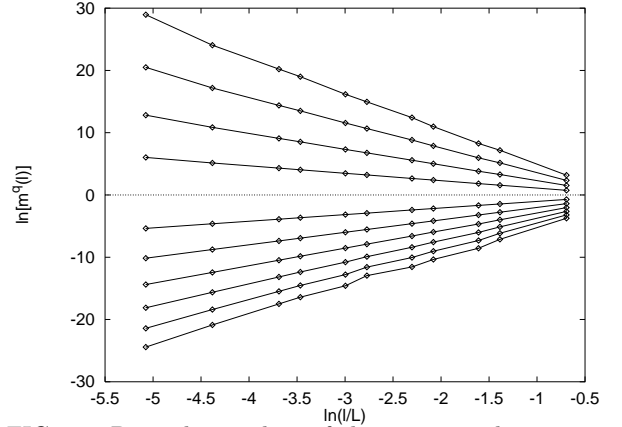,width=0.45\textwidth}
  \caption{Power-law scaling of the average $q$th-moment 
     of box masses $m_(l)$ for $q$-values from -2.0 to 3.0.}
    \label{fig:scale}
  \end{center}
\end{figure}

\begin{figure}[htbp]
  \begin{center}
    \leavevmode
    \epsfig{figure=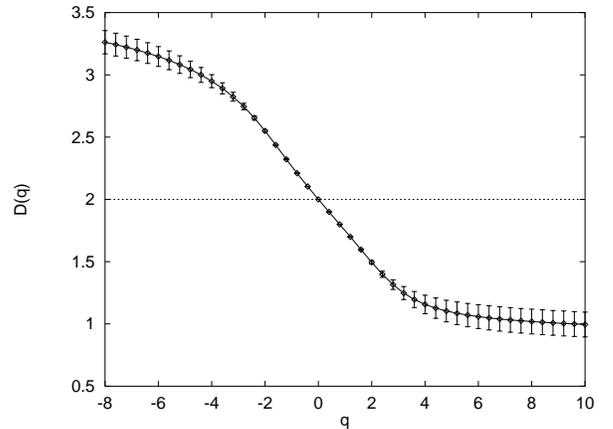,width=0.45\textwidth}
  \caption{The function $D(q)$ averaged over 12 wave functions.}
    \label{fig:Dq}
  \end{center}
\end{figure}

Fig.~\ref{fig:wave} shows the square-amplitude $\rho(r) = |\Psi(r)|^2$ 
of a typical eigenstate obtained in the Chalker-Coddington
network at critical energy $E=E_c=0$. Like the local spectral density
(see Sec. \ref{subsec-ldos}), the eigenstate exhibits a
self-similar, multifractal structure, generic for the
localization-delocalization transition point.
\begin{figure}[htbp]
  \begin{center}
    \epsfxsize 8cm
    \caption[Critical wave functions.]{
     Squared amplitude of a Critical wave functions 
     in a Chalker-Coddington network of 256x256 saddle points. Darker
     areas denote lower square amplitude.} 
    \label{fig:wave}
  \end{center}
\end{figure}
Similar as for the local spectral density, 
the scale invariance of the eigenfunctions can be 
demonstrated  by investigating the scaling behavior of the
quantity
\begin{equation}
  \label{box-prob}
  m(l)= \int_{l^2} d^2r \rho(r_0+r),
\end{equation}
which is the mass of a square of linear dimension $l$
with respect to the density $\rho(r)$.
The disorder averaged $q$th-moments of it, $ m_q(l)=\langle m^q(l) \rangle$,
obeys power law behavior over a wide range of box-sizes $l$, as can be
seen in the double logarithmic plot in Fig. \ref{fig:scale}.
The data were taken from the wave function in Fig. \ref{fig:wave}.
The generalized dimensions $D_q$ of the critical density $\rho(r)$, 
defined by $m_q(l) \propto l^{d+(q-1)D_q}$ ($d=2$), are plotted in 
Fig. \ref{fig:Dq}. For the determination of these exponents we used 
networks with sizes varying from $80\times 80$ up to $512 \times 512$
and of two different kinds of boundary conditions, full-periodic and
semi-periodic ones. To calculate eigenfunctions we applied the method
of inverse iteration and, to treat larger systems, the Lanczos algorithm.
Significant deviations in the multifractal exponents for networks 
of different sizes and boundary condition  were not found. 
The results are also independent on the used numerical method.
Table \ref{tab:expo}
lists some characteristic exponents and gives a comparison to 
the results obtained in the two models mentioned above. The exponent 
$\alpha_0$ is defined as the derivative of $D_q$ with respect to $q$
at $q=0$. The values for the
multifractal exponents of these models match within the errors.
\begin{table}[htb]
  \begin{tabular}{c|c|c|c|c|c} 
    Model & $ D_1 $&$ D_2 $&$ D_{-\infty} $&$ D_\infty $&$\alpha_0$\\ 
    \hline 
    N & $ 1.75 \pm 0.01 $&$ 1.50\pm0.01 $&$3.4\pm0.2 $&$ 1.1 \pm 0.1$&$
    2.26 \pm 0.01$\\ 
    \hline 
    TB & - &$ 1.62 \pm 0.02 $&$ 3.7\pm 0.2$&$ 1.0\pm 0.1$& 
    $2.29 \pm 0.02$ \\
    \hline
    RL &- &$ 1.5\pm 0.1 $&$3.7\pm 0.02 $&$0.95 \pm 0.2 $&
    $2.3\pm 0.07  $\\
    \hline
    \hline
    $D/\tilde D$&$2.01\pm 0.02 $&$ 2.00\pm 0.02 $&$ 2.3\pm 0.4 $&$ 1.8
    \pm 0.4$&$ 
    1.98 \pm 0.02$\\ 
  \end{tabular}
  \caption{Comparison of characteristic multifractal exponents for
    different models: network (N), tight-binding (TB) and
    the Random Landau (RL) model . The last row gives the ratio of
    corresponding exponents for critical eigenstates and critical
    local spectral density.} \label{tab:expo}
\end{table}
The corresponding exponents of the critical local density of states 
(with tilde) appear to be half of those of the critical states. This simple 
proportionality, $D_q = d \tilde D_q$, where $d=2$ is the spatial
dimension, is a consequence of dynamical scaling introducing a single
energy dependent length scale $L_\omega \propto \omega^{-1/d}$
\cite{huckestein-klesse}.

\section{Two-Point-Conductivities} 
\label{sec-pconduct}
\begin{figure}[htbp]
  \begin{center}
    \leavevmode
    \epsfig{figure=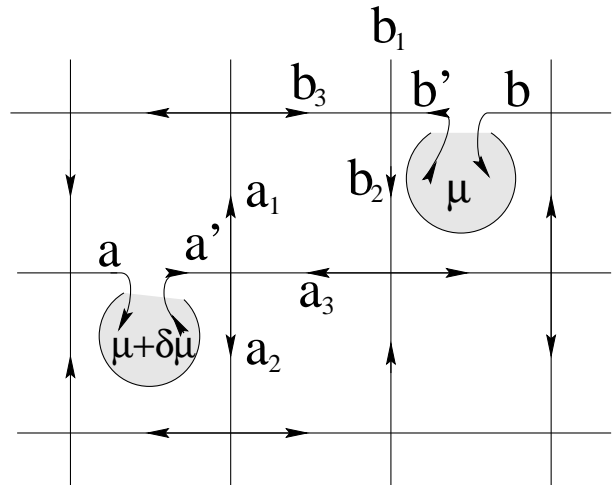,width=0.45\textwidth}
  \caption{Chalker-Coddington network coupled to external reservoirs 
           by two point contacts.}
    \label{fig-pcontact}
  \end{center}
\end{figure}

Finally, we come to an example of an open network.
We will study a Chalker-Coddington network
coupled to external reservoirs via ohmic contacts, in order to
demonstrate how the network model can be used for the determination
of conductivities. For simplicity we assume the the coupling to the
reservoirs is given by two ideal point contacts: two links, $a$ and
$b$, of a closed network in a distance $r$ are cut and connected to
external reservoirs of chemical potential $\mu_a, \mu_b$, respectively
see Fig. \ref{fig-pcontact}. This means that, firstly, the incoming 
channels $a'$ and $b'$ are occupied up to the chemical potentials 
$\mu_a,\mu_b$, secondly, the 
currents in the outgoing channels $a$ and $b$ are ideally absorbed by
the reservoirs.
 The goal is to calculate the (dimensionless) conductance $g$ of this
configuration obeying the equation $\delta I = e/h\: g \delta \mu$,
where $\delta \mu = \mu_a-\mu_b$ and $\delta I$ the net current
flowing through the network from reservoir $a$ to $b$. Here the entire
network can be viewed as a single scattering center -- with quite
complex internal structure -- connecting the two incoming channels
($a',b'$) to the outgoing ones ($a,b$). So, according to the
Landauer--B\"uttiker formula, the conductance $g$ is in this case the 
corresponding scattering amplitude $ |t_{ab}|^2$ from $a'$ to $b$,
which to determine is our task. To do this, consider the following
situation: let the incoming channel $a'$ be feed by a unit current 
with amplitude $1$ and set the amplitude in the other one, $b'$, 
to $0$. The current will distribute over the network obeying (i)
the boundary condition $\psi_{a'}=1$, $\psi_{b'}=0$ at the external
incoming links $a'$,$b'$ and (ii) the local scattering conditions
$\psi_l = \sum_{m} t_{lm} \psi_m$ at internal links $l$ (including 
$a$ and $b$). Then, $\psi_b$, the amplitude in outgoing channel $b$,
equals exactly $t_{ab}$, the coefficient we are looking for.
 It is straightforward to show that conditions (i) and
(ii) are equivalent to the vector equation
\ben\label{vector-equation}
U(\Psi-\psi_a\bfe_a + \bfe_a - \psi_b \bfe_b) = \Psi,
\een
where all quantities refer to the {\em closed} network. The term 
$-\psi_a\bfe_a$ belongs to the current reflected back into reservoir
$a$, $+\bfe_a$ corresponds to current fed into channel $a'$, and
$-\psi_b \bfe_b$ is due to current out-flowing through channel $b'$.
With projection operators $\bar P_l = 1-P_l = 1 - |\bfe_l><\bfe_l|$
Eq. \Ref{vector-equation} reduces to 
\be
(1-U\bar P_a \bar P_b)\Psi = U \bfe_a,
\ee
and we obtain finally,
\bea
t_{ab} &\equiv& \psi_b \equiv \langle \bfe_b, \Psi \rangle \\
&=& \langle \bfe_b, (1-U\bar P_a\bar P_b)^{-1}U \bfe_a \rangle.
\eea
By this, the problem of finding the two-point conductance is reduced
to a standard problem of linear algebra, which we solved by numerical 
LU-decomposition. In the following we will present data obtained by
this method for critical Chalker-Coddington networks.

At criticality the conductances are expected to show a power law
behavior with  
respect to the distance $r\ll L$ between the contacts,
while their distribution will be very broad for any $r$. Furthermore,
as was pointed out by Zirnbauer \cite{zirnbauer}, there should not be
a dependence of the conductance on the system size.
As we will
see, such a picture is consistent with our data.
We have determined conductances for systems of size $L=40,60,100$ and
distances between links varying from $r=1$ to $r=L/2$. For every distance
we determined up to 1300 conductances (depending on the system size).

\begin{figure}[hbtp]
  \begin{center}
    \leavevmode
    \epsfig{figure=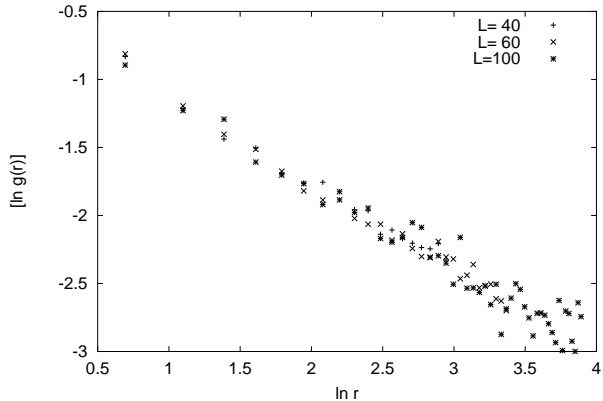,width=0.45\textwidth}
  \caption{The logarithm of the typical conductance plotted
    against $\ln r$ for three different system sizes.}
    \label{fig:curve}
  \end{center}
\end{figure}

The broadness of the distribution suggests to determine the moments
$[g^q]$, the geometric mean $g_t = \exp[\ln g]$ and the log-variance
$\left[(\ln g/g_t)^2\right]$. 
In Fig.~\ref{fig:curve} we show $[\ln g]$ plotted against 
$\ln r$ for three different system
sizes $L=40,60,100$. There is no visible $L$ dependence, as expected,
and the typical value scales in a power law fashion with respect to $r$.

\begin{figure}[htbp]
  \begin{center}
    \leavevmode
    \epsfig{figure=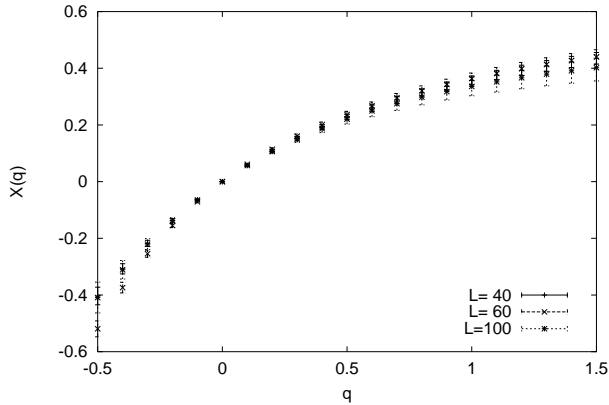,width=0.45\textwidth}
  \caption{The function $X(q)$ for three different system sizes.}
    \label{fig:X}
  \end{center}
\end{figure}

We find similar behavior for the moments $[g^q]$, whose behavior can
be described in the following way:
\begin{equation}
  \label{codb}
 [g^q]_L\propto r^{-X(q)},
\end{equation}
where $X(q)$ is called the multifractal spectrum of the conductance.
The numerical data for $X(q)$ for system sizes 40, 60 and 100 
is depicted in Fig.~\ref{fig:X}.
Within the error bars, the non-linear curves for different 
system sizes match, indicating universal multifractality.
For further information on the distribution $P(g)$ of the
point-contact conductances see ref.~\cite{neues_paper}.


\section{Summary}
We extended the Chalker-Coddington model and a 3D 
generalization of it by a simple, discrete time evolution and
studied dynamical and spectral aspects of the resulting dynamical 
network models. Despite of the simplicity of the dynamics, the models
correctly show the universal aspects of normal and critical
diffusion as well as of Anderson localization in temporal evolution
of network wave functions. In case of closed networks, the analogue
of eigenfunctions and eigenenergies exist in form of eigenvectors 
and quasi-energies (=eigenphases) of an unitary network operator.
We investigated spatial and spectral correlation of eigenstates and
quasi-energy spectra and found qualitative and quantitative agreement
with theoretical predictions and other numerical work within 
conventional models. Furthermore, we used an open network for
the calculation of two-point conductances.

Definitions and methods with regard to the
network dynamics are not specific to the
two models investigated in particular. Therefore, they should be
as well applicable to any other model formulated as a network 
in terms of local scatterers and connecting link-channels. 

Disregarding physical and mathematical subtleties, the analogies
of a physical system described by a Hamiltonian $H$ with
it's corresponding network model represented by an unitary operator
$U$ can be summarized as in Tab. \ref{tab:relation}.
\begin{table}[htb]
  \begin{tabular}{c|c}                 $e^{-iH\tau}$ & $U$ \\
    \hline 
    $\Psi(t) = e^{-iHt}\Psi(0) $ & $\Psi(k\tau) = U^k \Psi(0) $ \\
    \hline
    $ H\psi_n = \eps_n \psi_n$ &  $U\Psi_n = e^{i\omega_n} \Psi_n$     \\
    \hline 
    $ \dots \eps_{n-1}, \eps_n, \eps_{n+1},     \dots$ &
    $\dots \omega_{n-1}, \omega_n, \omega_{n+1},  \dots $       \\
  \end{tabular}
  \caption{Relation between Hamiltonian and network operator
    (schematic). }
  \label{tab:relation}
\end{table}
Notice that there are still significant differences: first of all, 
the dimensions of $H$ and $U$ do not match. The dimension of $U$
equals the number of links, which generally is much smaller than
the dimension of $H$, since each link-channel itself contains 
a large number of (quasi-) asymptotic states. In some sense, 
changing from $H$ to $U$ corresponds to reducing the original 
system by non-interesting degrees of freedom. In case of the 
models investigated here, these degrees of freedom are the 
link states: the network wave function $\Psi$ no longer contains any
information about their structure (e.g. wave number). Obviously, 
this dimensional reduction constitutes the main advantage of the network
model, especially in numerical applications. The price to be
paid for that is a loss of accuracy. Processes and correlations
corresponding to length and energy scales of the out-projected
internal degrees of freedom can no longer be reproduced. In particular
this means that the statistical equivalence of energy and quasi-energy
spectrum holds only in the vicinity of a certain energy $E_0$ to which
$U=U(E_0)$ refers. Similarly, the network time evolution can only be 
related to the time evolution of real wave packets composed of 
eigenstates with energies close to $E_0$.



\acknowledgments
Fruitful discussions with J\'anos Hajdu, Martin Janssen and Bodo
Huckestein are gratefully acknowledged. 
This work was performed within the research program of the
Sonderforschungsbereich 341 of the Deutsche Forschungsgemeinschaft.
M.M. thanks the Deutsche Forschungsgemeinschaft, R.K
the Minerva foundation (Munich, Germany) and the Weizmann Institute
of Science (Israel) for support.


\end{document}